\documentclass{mem}
\usepackage{natbib}
\usepackage{txfonts}
\usepackage{graphicx}
\idline{1}{1}
\begin{document}

\title{
Red Variables in Globular Clusters
}

   \subtitle{Comparison with the Bulge and the LMC}

\author{
N. \, Matsunaga\inst{1} 
\and Y. \, Nakada\inst{2,1}
\and T. \, Tanab\'e\inst{1}
\and H. \, Fukushi\inst{1}
\and Y. \, Ita\inst{3}
          }

  \offprints{N. Matsunaga (\email{matsunaga@ioa.s.u-tokyo.ac.jp})}

\institute{
Institute of Astronomy, School of Science, The University of Tokyo,
2-21-1 Osawa, Mitaka, Tokyo 181-0015, Japan
\and
Kiso Observatory, School of Science, The University of Tokyo,
Mitake, Kiso, Nagano 397-0101, Japan
\and
Institute of Space and Astronomical Science, Japan Aerospace
Exploration Agency, 3-1-1 Yoshinodai, Sagamihara, Kanagawa 229-8510, Japan
}

\authorrunning{Matsunaga et al.}

\titlerunning{Red Variables in Globular Clusters}

\abstract{
We are conducting a project aimed at surveys and repeated observations
of red variables (or long-period variables) in globular clusters.
Using the IRSF/SIRIUS near-infrared facility located at South Africa,
we are observing 145 globular clusters that are accessible
from the site.
In this contribution, we present our observations and preliminary results.
We have discovered many red variables, especially in the Bulge region,
whose memberships to the clusters remain to be confirmed.
Using a sample of all red variables
(both already known and newly discovered ones) in globular clusters
except those projected to the Bulge region,
we produce a $\log P$-$K$ diagram and compare it
with those for the Bulge and the Large Magellanic Cloud.
A prominent feature is that the bright part of overtone-pulsators' sequence
($B^+$ and $C^\prime$) is absent.
We discuss its implication on the evolution of red variables.
}

\maketitle{}

\section{Introduction}
The study on red variables has advanced very much since \citet{Wood-1998}
found parallel sequences on the period-magnitude diagram
as a result of the MACHO project for the Large Magellanic Cloud.
Many papers have been published on the variables in the Magellanic Clouds
(Ita et al. 2004a, 2004b; Kiss \& Bedding, 2003, 2004;
also see the contribution by Dr. Kiss in this volume).
However, it is difficult to discuss the stellar evolution directly
because the Magellanic Clouds contain red variables with different ages
and different chemical components.

In contrast, globular clusters (GCs) are systems of
single stellar population, so that we can tell the age and
the initial chemical component for each variable star.
Making use of this advantage, some authors have worked on
the evolution of red variables in GCs.
\citet{Whitelock-1986} found
Mira variables and semi-regular variables in GCs
define a relatively tight sequence on the $\log P$-$M_{\rm bol}$ diagram.
\citet{Bedding-1998} considered the sequence as an evolutionary track,
whose slope is consistent with an evolutionary calculation
by \citet{Vassiliadis-1993}.
On the other hand, \citet{Perl-1990} theoretically examined
the $\log P$-$M_{\rm bol}$ diagram with a staircase pattern
for variables in 47~Tuc, and concluded that
a switch of pulsational-mode occurs at a certain luminosity.
Recently, \citet{Lebzelter-2005} also proposed the switch of mode
based on $\log P$-$K$ diagram and
their measurements of variations of radial velocities.

These works clearly demonstrated that GCs are
good testbeds for studying the evolution of red variables.
However, early surveys were not complete, and many of them were optimized
to search for short-period variables such as RR Lyrae variables
\citep{Clement-2001}.
In addition, they are mostly conducted in visible wavelength.
Some red variables are undergoing heavy mass loss and
holding thick circumstellar materials, which make the stars only
detectable in infrared\citep{Tanabe-1997}.
So that we decided to conduct a near-infrared survey for red variables
in GCs.

\section{Observations and Preliminary Results}
We started our near-infrared survey in the spring of 2002, using
the IRSF {1.4~m} telescope and the SIRIUS camera. They are
established by Nagoya University and National Astronomical Observatory
of Japan, and sited at Sutherland station of
South African Astronomical Observatory. They can take
three images in $J$-, $H$-, and $K_{\rm s}$-band simultaneously
with the $8^\prime \times 8^\prime$ fields-of-view.
For the details of the IRSF and the SIRIUS,
see \citet{Nagashima-1999} and \citet{Nagayama-2003}.

There are 150 GCs found in the Milky Way and the neighborhood
\citep{Harris-1996}.
From the IRSF site, we can reach to about $+30^\circ$ in Declination,
and we have observed such 145 clusters at least once.
Until the end of 2004, we have observed 133 clusters more than 10 times
and 40 clusters more than 20 times.

With data analyses for the 133 clusters, we detected variations of
58 known red variables in 30 clusters and discovered
185 red variables in 44 clusters.
Detection limits of variability is typically better than 0.1 mag.
Many of the newly discovered variables lie in the direction of
the Galactic Bulge. 31 clusters with the new variables
locate within the $|l|<10^\circ$ and $|b|<10^\circ$ region.
Because the Bulge itself contains many
red variables, it is important to check their memberships.

Especially, it is urgent for some peculiar objects. 
Among the new ones, there are very red stars, e.g. $(J-K)_0 > 3$.
It is not known that such stars exist in GCs.
We detected SiO maser emissions, evidences of relatively thick
circumstellar materials, from four of the newly discovered
variables and also from two known variables \citep{Matsunaga-2005a}.
The velocities of the emissions indicate five of them
can be associated with GCs.

Using the near-infrared surface brightness obtained by the {\it COBE}/DIRBE
data \citep{Matsunaga-2005b},
roughly 20 \% of the detected variables are considered to be
associated with the GCs.

\section{Discussions}

\begin{figure*}[]
\resizebox{\hsize}{!}{\includegraphics[clip=true]{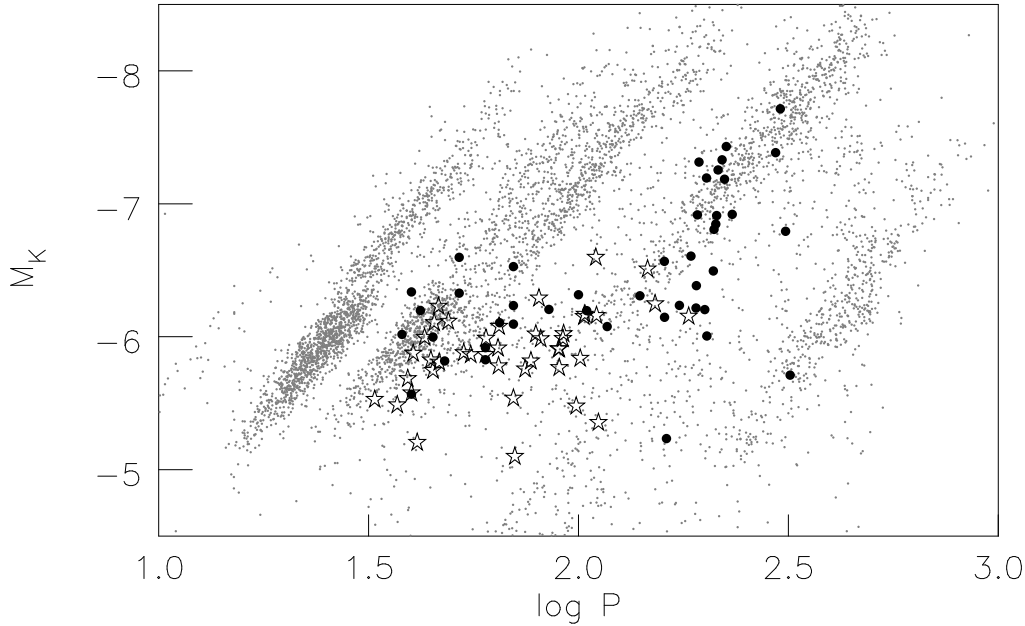}}
\caption{
\footnotesize
The $\log P$-$K$ diagram. Filled circles for variables in metal-rich
clusters (${\rm [Fe/H]} > -1$), star symbols for those in metal-poor ones,
and gray points for the data of the Large Magellanic Cloud \citep{Ita-2004b}.
}
\label{fig:PLR}
\end{figure*}

\subsection{The $\log P$-$K$ diagram and comparison with the LMC}
In the following, we will discuss the $\log P$-$K$ diagram using our samples.
In order to avoid the Bulge field contamination,
we adopt variables whose galactic coordinates are
out of the Bulge range: $|l|>10^\circ$ and/or $|b|>10^\circ$.
Collecting red variables with known periods and near-infrared measurements,
our sample contains 86 red variables in 29 GCs.
For 27 stars among them, we detected their variations and obtained
their mean magnitudes (mean of minimum and maximum);
we did not detected significant variations for 50 stars;
9 stars are located too north
to be observed and we just adopted the magnitudes listed in
the 2MASS point source catalog.

Fig. \ref{fig:PLR} shows the $\log P$-$K$ diagram.
Filled circles for variables in metal-rich GCs ([Fe/H]$>-1$) and
triangles for those in metal-poor ones are superposed on
gray points for those in the Large Magellanic Cloud\citep{Ita-2004b}.
Mira variables in the metal-rich clusters obey the sequence $C$, and
some variables are located at around the sequence $B^-$
(notations for the sequences are based on those of \citealt{Ita-2004a}).
As is well known, bright Mira variables belong to only metal-rich
GCs \citep{Frogel-1998}.

Few variables in GCs lie on the sequences $A$, $B^+$,
$C^\prime$, and $D$. For variables on the sequences $A$ and $D$,
the periods are difficult to get. Those on the sequence $A$
have small amplitude and irregular light curves,
and ones on the sequence D have very long periods and sometimes
double-periodic light curves \citep{Wood-2000}.
We need massive photometric data and very careful analyses in order to
conclude the absence of these sequences. 

On the other hand, variables on the sequences $B^+$ and $C^\prime$
should be easier
to find, at least compared with those on the sequence $B^-$.
It is also known about GCs that all the stars brighter than the tip of
Red Giant Branch (RGB) are found to be Mira variables
(see \citealt{Frogel-1998}, for example).
So that, we conclude this absence is a real feature of variables in GCs.
Please note that this absence was utilized to propose
the evolutionary track by \citet{Perl-1990} and \citet{Lebzelter-2005}.
We confirmed the absence from a larger sample including many GCs.

\subsection{Comparison with the Bulge and its implication\label{sec:Bulge}}

We found the intermediate part of the sequence $C^\prime$(or $B^+$)
exists on the $\log P$-$K$ diagram for the Galactic Bulge
given by Glass \& Schultheis (2003; see their Fig. 7).
On that sequence, there is an vacant region at the brightest part,
but some variables are as bright as $M_K=-7.5$ with the distance modulus
assumed to be 14.5~mag. Although it is difficult to distinguish between
the sequences $C^\prime$ and $B^+$,
they are bright overtone pulsators that are not found in GCs.
The top of the sequence is brighter than the tip of RGB.
Although there are some uncertainties
about the tip of RGB for the Bulge,
such as spatial depth and metallicity effects,
we consider they should work on both the sequences A and B.
We conclude the intermediate part is
in the Asymptotic Giant Branch (AGB) phase.

It is suggested that AGB stars in the Bulge take the evolutionary
track with the staircase pattern similar to that of 47~Tuc,
but the luminosity at which the mode of pulsation switches is different.
This can be interpreted that the variables in the Bulge have
larger initial masses than those in 47~Tuc and take
a brighter evolutionary track as presented by \citet{Vassiliadis-1993}.
The star formation history of the Bulge is still controversial,
but the majority of the constituent stars
have the similar ages to those of GCs\citep{Zoccali-2003}.
Nonetheless, their high metallicities have an effect on initial masses
of the red variables.
Theoretically, red giants with the high initial metallicities are expected
to have the larger initial masses \citep{Iben-1984}.

It is important to discuss GCs in the Bulge region
whose metallicities are similar to the Bulge field population
after confirming the memberships of variable stars.
It is also interesting to investigate the $\log P$-$K$ diagram for
some massive clusters with the intermediate-age
($\sim$1~Gyr, i.e. more massive) AGB stars
in the Magellanic Clouds.

%It is interpreted that some of these variables are on the Red Giant Branch
%(RGB; or First-ascent Branch),
%and their luminosities depend on the metallicity so that more metal-rich
%clusters have brighter tips of RGB \citep{Ferraro-2000}.

\section{Conclusions}
We discovered many red variables
in the course of our near-infrared survey toward 145 GCs.
Many of the newly discovered variables lie in the direction of
the Galactic Bulge. Since there is a large field population of
red variables in the Bulge, it is important to check their memberships
by radial velocities and proper motions.

Using the sample outside the Bulge region including both the new variables
and already-known ones (86 variables in 29 clusters),
we discussed the distribution on the $\log P$-$K$ diagram.
We found the absence of the bright part of the sequences $C^\prime$ and $B^+$,
which was known for 47~Tuc, with more variables whose metallicities are
widely spread. We also compared the distribution with that of the Bulge.
It is suggested that AGB variables switch from
overtone pulsators to Mira variables at a certain luminosity,
which depends on stellar parameters such as initial mass and metallicity.

\begin{acknowledgements}
We are grateful to the IRSF/SIRIUS team for the well-operating instrument
and their supports for our observation. Parts of our observations are
kindly conducted by {Dr. M. Feast}, {Dr. J. Menzies}, {Dr. P. Whitelock},
{Dr. E. Olivier}, {Dr. T. Nagata}, {Dr. H. Nakaya}, {Dr. T. Naoi},
{Mr. S. Nishiyama}, {Mr. D. Baba}, and {Ms. A. Ishihara}.
One of the authors (N. Matsunaga) is financially supported by
the Japan Society for the Promotion of Science (JSPS) for Young Scientists.
\end{acknowledgements}

\bibliographystyle{aa}

\begin{thebibliography}{}

\bibitem[{Bedding \& Zijlstra (1998)}]{Bedding-1998}
Bedding, T.~R., Zijlstra, A.~A., 1998, \apj, 506, L47
\bibitem[{Clement et al. (2001)}]{Clement-2001}
Clement, C.~M., et al. 2001, \aj, 122, 2587
%\bibitem[{Ferraro et al. (2000)}]{Ferraro-2000}
%Ferraro, F.~R., Montegriffo, P., Origlia, L., Fusi Pecci, F.,
%2000, \aj, 119, 1282
\bibitem[{Frogel \& Whitelock (1998)}]{Frogel-1998}
Frogel, J.~A., Whitelock, P.~A., 1998, \aj, 116, 754
\bibitem[{Harris (1996)}]{Harris-1996}
Harris, W.~E., 1996, AJ, 112, 1487
\bibitem[{Glass \& Schultheis (2003)}]{Glass-2003}
Glass, I. S., Schultheis, M., 2003, MNRAS, 345, 39
\bibitem[{Iben \& Renzini (1984)}]{Iben-1984}
Iben, I., Renzini, A., 1984, Phys. Rep. 105, 329
\bibitem[{Ita et al. (2004a)}]{Ita-2004a}
Ita, Y. et al. 2004a, \mnras, 347, 720
\bibitem[{Ita et al. (2004b)}]{Ita-2004b}
Ita, Y. et al. 2004b, \mnras, 353, 705
\bibitem[{Kiss \& Bedding, (2003)}]{Kiss-2003}
Kiss, L.~L., Bedding, T.~R. 2003, \mnras, 343, L79
\bibitem[{Kiss \& Bedding, (2004)}]{Kiss-2004}
Kiss, L.~L., Bedding, T.~R. 2004, \mnras, 347, L83
\bibitem[{Lebzelter et al. (2005)}]{Lebzelter-2005} 
Lebzelter, T., et al., 2005, \aap, 432, 207
\bibitem[{Matsunaga et al. (2005a)}]{Matsunaga-2005a} 
Matsunaga, N. et al. 2005, \pasj, 57, L1
\bibitem[{Matsunaga et al. (2005b)}]{Matsunaga-2005b} 
Matsunaga, N., Fukushi, H., Nakada, 2005, \mnras, in press
\bibitem[{Nagashima et~al. (1999)}]{Nagashima-1999}
Nagashima, C., et~al. 1999,
  in Proceedings of Star Formation 1999, Star Formation, ed. T.~Nakamoto
  (Nagoya, Japan), 397
\bibitem[{Nagayama et~al. (2003)}]{Nagayama-2003}
Nagayama, T., Nagashima, C., Nakajima, Y., {et~al.} 2003, in SPIE Vol. 4841,
Instrument Design and Performance for Optical/Infrared Ground-based Telescopes,
ed. M.~Iye \& A.~F.~M. Moorwood, 459
\bibitem[{Perl \& Tuchman (1990)}]{Perl-1990}
Perl, M., Tuchman, Y., 1990, \apj, 360, 554
\bibitem[{Tanab\'e et al. (1997)}]{Tanabe-1997}
Tanab\'e, T., et al. 1997, Nature, 385, 509
\bibitem[{Vassiliadis \& Wood (1993)}]{Vassiliadis-1993}
Vassiliadis, E., Wood, P.~R., 1993, ApJ, 413, 641
\bibitem[{Whitelock (1986)}]{Whitelock-1986}
Whitelock, P. A., 1986, \mnras, 219, 525
\bibitem[{Wood (2000)}]{Wood-2000}
Wood, P.~R., 2000, PASA, 17, 18
\bibitem[{Wood et al. (1998)}]{Wood-1998}
Wood P.~R., Habing, H.~J., McGregor, P.~J., 1998, \aap, 336, 925
%\bibitem[{Wood \& Sebo (1996)}]{Wood-1996}
%Wood, P.~R., Sebo, K.~M., 1996, \mnras, 282, 958
\bibitem[Zoccali et al.(2003)]{Zoccali-2003}
Zoccali, M., et al. 2003, \aap, 399, 931

\end{thebibliography}

\end{document}